\pdfoutput=1
\documentclass[twocolumn,showpacs,preprintnumbers,amsmath,amssymb]{revtex4}

\def\publishtype{electronic}
\long\def\beginpgfgraphicnamed#1#2\endpgfgraphicnamed{\includegraphics{#1}}

\usepackage[pdfborder={0 0 0}]{hyperref}

\usepackage{graphicx}
\usepackage{dcolumn}
\usepackage{bm}
\usepackage{SIunits}
\usepackage{ifthen}

\usepackage{movie15}

\newcommand {\be}{\begin{equation}}
\newcommand {\ee}{\end{equation}}
\newcommand {\bea}{\begin{eqnarray}}
\newcommand {\eea}{\end{eqnarray}}
\newcommand {\bem}{\begin{displaymath}}
\newcommand {\eem}{\end{displaymath}}
\newcommand {\f}{\frac}
\newcommand {\p}{\partial}
\newcommand\D[2]{\frac{\partial #1}{\partial #2}}
\def\to{\;\text{to}\;}

\begin{document}

\preprint{ }
\title{The effects of superconductor-stabilizer interfacial resistance on quench of current-carrying coated conductor}
\author{ G. A. Levin} 
\affiliation{Air Force Research Laboratory, Propulsion Directorate, Wright-Patterson Air Force Base, OH 45433}
\author{ K. A. Novak}
\affiliation{Department of Mathematics, Air Force Institute of Technology, Wright-Patterson Air Force Base, OH 45433}
\author{P. N. Barnes}
\affiliation{Air Force Research Laboratory, Propulsion Directorate, Wright-Patterson Air Force Base, OH 45433}

\date{\today}

\begin{abstract}
We present the results of numerical analysis of a model of normal zone propagation in coated conductors. The main emphasis is on the effects of increased contact resistance between the superconducting film and the stabilizer on the speed of normal zone propagation, the maximum temperature rise inside the normal zone, and the stability margins. We show that with increasing contact resistance the speed of normal zone propagation increases, the maximum temperature inside the normal zone decreases, and stability margins shrink. This may have an overall beneficial effect on quench protection quality of coated conductors. We also briefly discuss the propagation of solitons and development of the temperature modulation along the wire.
\end{abstract}
\maketitle

\section{\label{sec:introduction}Introduction\protect}

Quench protection of large scale devices, such as magnets and cables, based on coated conductors has emerged as one of the major unresolved obstacles in their implementation. High operating temperature and, correspondingly, relatively large heat capacity make coated conductors very stable in comparison to the conventional low temperature superconductors. However, the side effect of this positive quality is that when a normal zone does nucleate it expands very slowly. The potential drop across a short normal section of a long conductor is difficult to detect and in adiabatic or nearly adiabatic conditions the temperature of this section may rise above the safe limit resulting in irreversible damage to the whole coil or cable strand. 

This article presents the results of a numerical analysis of a model of normal zone propagation (NZP) specialized to the architecture of the state-of-the-art coated conductors. Its main purpose is to elucidate the effects of the interfacial resistance (contact resistance) between the superconducting YBa$_2$Cu$_3$O$_{7-x}$ (YBCO) film and copper stabilizer on stability and speed of NZP{}. The interest to this problem arose initially from an effort to understand some peculiar effects that accompany quench in coated conductors~\cite{Duckworth,Wang}. It seems clear now that these phenomena result from a large resistance between the YBCO film and a metal substrate~\cite{Breschi,Levin1}. This understanding has lead to realization that increasing the contact resistance between the YBCO film and copper stabilizer may have beneficial effect on the speed of the normal zone propagation~\cite{ASC,Wan,PRE}. 

The effects of a large contact resistance between the stabilizer and conventional low $T_c$ superconductors have been studied extensively in the past~\cite{Ah1,A,G,Ah2,Gurevich}. However, the idea of tailoring the properties of the superconducting wires by increasing the contact resistance has not been adopted to wider use. This option of conductor design had lain dormant for many years---a solution in wait of a problem. Perhaps, coated conductors present just such a problem. Increasing the contact resistance does make the conductor less stable. However, since coated conductors are inherently much more stable than the low $T_c$ superconductors, the reduction of the stability margins accompanied by increasing the speed of NZP may allow to develop coated conductors overall better suited for large scale applications than their current version with a minimized contact resistance. 

Here we will discuss the NZP in a straight coated conductor cooled from the surface. This model more closely describes the typical conditions in the experiments such as~\cite{Duckworth,Wang} or in a  superconducting cable, rather than in a pancake coil. Correspondingly, the concrete example presented below is based on the operating temperature equal to \unit{65}{\kelvin} and the values of the material parameters in the temperature range $\unit{65 \to 77}{\kelvin}$. Our main conclusions that the increased contact resistance leads to increased NZP speed and reduced stability margins will remain qualitatively valid for any type of application. However, in order to adequately  describe a pancake coil one needs to take into account the heat transfer in the radial direction, between the turns, in addition to the lateral heat flux which we consider here. The problem of NZP in a pancake coil will be addressed elsewhere. 

This paper is structured as follows. In Section II we briefly formulate the model of a quench in coated conductor based on conditions of energy and charge conservation. The model is similar to that in~\cite{ASC,PRE}, but with a modified constituent relationship between the electric field and current density in the superconductor.  However, in~\cite{ASC,PRE} the numerical solutions were obtained using an approximation that is strictly valid only in the limit of a small contact resistance (see detailed explanation in~\cite{PRE}). Therefore, the results of that analysis could not be extended to the conditions when the contact resistance is arbitrarily large. Moreover, legitimate questions remained as to what extent the results of~\cite{ASC,PRE} were influenced by the approximation used to solve the problem, rather than the physics of the phenomenon. The solutions of the system of coupled nonlinear partial differential equations were obtained rigorously without approximations. Qualitatively, the current results are similar to those in~\cite{ASC}. The propagation speed increases and stability margins decrease with increasing contact resistance. Quantitatively however, there are substantial differences in the rate of change of these characteristics with the value of contact resistance. The rigorous solutions also confirm the emergence of the dissipative structures---spontaneous modulation of temperature along the conductor---reported previously~\cite{PRE}.

\section{\label{sec:model}Model\protect}
\begin{figure}[ht]
\centering
\beginpgfgraphicnamed{fig-sketch}
\input{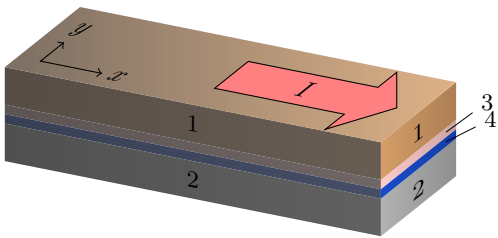}
\endpgfgraphicnamed
	\caption{\label{fig:sketch}%
A sketch of the cross-section of coated conductor (not to scale). Shown are four constituencies we have taken into account in this model. These are copper stabilizer (1), metal substrate (2), a thin superconducting film deposited on top of the substrate (4), and the interfacial resistive layer (contact resistance) that segregates the superconducting film from the stabilizer (3).}
\end{figure}
Coated conductors~\cite{Larbalestier,Foltyn} are manufactured in the form of a tape in which the superconducting YBCO film of about \unit{1}{\micro\meter} thick is deposited on a buffered flexible metal substrate (e.g. Ni-W alloy, Hastelloy or stainless steel). A copper stabilizer is either soldered or electroplated on top of the YBCO film.  See Figure~\ref{fig:sketch}. The standard width of such a tape-like wire is \unit{4}{\milli\meter}, the thickness, about evenly divided between the substrate and stabilizer, is close to \unit{100}{\micro\meter}. 

In~\cite{PRE} it was shown that the 3D equation of heat conduction in a thin tape-like composite wire can be reduced to a 2D (planar) or 1D (linear) model if the heat flux from the surface does not greatly exceed \unit{1}{\watt\per\centi\meter\squared}. In this case the temperature variation across the thickness of the tape constitutes a fraction of a degree and is much smaller than the variation of temperature along the wire. The 1D approximation is valid as long as the thermal diffusion length is greater or comparable to the conductor width. In coated conductors this condition is met. The 1D (in-plane) heat conduction equation for the coated conductor has the form~\cite{PRE}
\begin{equation}\label{E:heat_equation_1}
C\f{\p T}{\p t}-\f{\p}{\p x}\left (K\f{\p T}{\p x}\right ) =Q -2K_0(T-T_0).
\end{equation}
Here $C=C_1d_1+C_2d_2$ is the combined heat capacity, $K=K_1d_1+K_2d_2$ is the effective thermal conductivity. The subscripts 1 and 2 refer to the stabilizer and substrate, respectively. The thicknesses of the stabilizer and substrate are denoted as $d_1$ and $d_2$. $Q=\int_{-d_2}^{d_1}q(z)dz$ is the density of the internal heat sources integrated over the thickness of the wire. $K_0$ is the heat transfer coefficient across the insulation on the surface of the wire and $T_0$ is the ambient temperature. 

The instantaneous redistribution of current between the superconducting film and stabilizer is determined by the condition of charge conservation 
\begin{equation}\label{E:charge_conservation_1}
\D{J_1}{x}=-\f{V_1-V_s}{\bar {R}},
\end{equation}
where $J_1$ is the linear density of current ($\ampere\per\centi\meter$) flowing through the stabilizer. $V_1$ and $V_s$ are the local electric potentials of the stabilizer and superconductor, respectively and $\bar {R}\;[\ohm\usk\centi\meter\squared]$ is the resistance of the unit area of the interface (contact resistance). This condition can also be used in the form
\begin{equation}\label{E:charge_conservation_2}
\D{}{x}\left (\bar{R}\D{J_1}{x}\right )= E_1 - E_s.
\end{equation}
Here $E_1=-\p V_1/\p x$ and $E_s=-\p V_s/\p x$ are the electric fields in the stabilizer and superconductor, respectively, and we do not assume that $\bar {R}$ is uniform.

The integrated area density of heat sources in~\eqref{E:heat_equation_1} takes the form
\begin{equation}\label{E:integrated_internal_heat_sources}
Q=\f{d_1}{\rho_1}E_1^2 +\f{(V_1-V_s)^2}{\bar {R}} 
+J_sE_s,
\end{equation}
where $J_s=J-J_1$ is the density of current in the superconductor and $J$ (constant) is the total transport current density in the coated conductor.

The constituent relationship for a superconductor can be presented in many forms. Here we will use the one from~\cite{Aronson}
\begin{equation}\label{E:SC_constituent_relationship}
E_s(J_s)= R_nJ_0\ln \left(1+\exp\{(J_s-J_c)/ J_0 \} \right ).
\end{equation}
Here $J_c(T)$ is the critical current and $R_n$ and $J_0$ are phenomenological parameters. They can be determined from the limiting cases. It is customary to define the critical current by the condition that at $J_s=J_c$ the electric field in the superconductor is equal to $E_0=\unit{1}{\micro\volt\per\centi\meter}$. Thus,
\begin{equation}
R_nJ_0=E_0/\ln 2.
\end{equation}
On the other hand, when $J_s$ substantially exceeds $J_c$ the electric field is determined by the flux flow
\begin{equation}
E_s(J_s)\sim \f{\rho_s}{d_s}(J_s-J_c),
\end{equation}
where $\rho_s$ is the normal state resistivity and $d_s$ the thickness of the superconducting film. Thus,
\begin{equation}
R_n\sim \f{\rho_s }{d_s}
\end{equation}
and
\begin{equation}
J_0\sim\f{E_0d_s}{\rho_s}.
\end{equation}
In coated conductors $d_s\sim \unit{1}{\micro\meter}$ and at $\unit{100}{\kelvin}$ the value of $\rho_s\sim \unit{100}{\micro\ohm\usk\centi\meter}$, so that 
\begin{equation}
J_0\sim \unit{10^{-6}}{\ampere\per\centi\meter}.
\end{equation} 
Thus, for all practical purposes~\eqref{E:SC_constituent_relationship} can be used in the piecewise form
\begin{equation}\label{E:piecewise_field}
E_s=\begin{cases}
R_n (J_s-J_c),&\text{if } J_s>J_c\\
0,&\text{if } J_s\le J_c
\end{cases}.
\end{equation}
Hereafter, we will adopt a linear temperature dependence  of $J_c$~\cite{Ekin,Gurevich}:
\begin{equation}\label{E:linear_temperature_dependence}
J_c=a(T_c-T);\;\; T<T_c.
\end{equation}
At $T>T_c$ the superconductor has ohmic resistance 
\begin{equation} 
E_s=\f{\rho_s}{d_s}J_s; 
\end{equation}
For the stabilizer, the conventional Ohmic relationship will suffice at all temperatures:
\begin{equation}
E_1(J_1)=\f{\rho_1}{d_1}J_1.
\end{equation}
In the normal state the resistance of YBCO film is much greater than that of the stabilizer,
\begin{equation}
\f{\rho_s}{d_s}\gg \f{\rho_1}{d_1}.
\end{equation}

The final step in formulating this model is to present~\eqref{E:heat_equation_1} and~\eqref{E:charge_conservation_2} in the dimensionless form. The current sharing temperature $T_1$ is defined by the condition $J_c(T_1)=J$.  Let us introduce a dimensionless temperature $\theta$ 
\begin{equation}
\theta=\f{T-T_1}{T_c-T_1}.
\end{equation}
Then,~\eqref{E:linear_temperature_dependence} takes the form
\begin{equation}\label{E:dimensionless_linear_temperature_dependence}
J_c=J(1-\theta).
\end{equation}
Let us introduce a fraction of the total current that flows through the stabilizer
\begin{equation}
J_1=Ju; \quad J_s=J(1-u); \quad 0\le u\le1.
\end{equation}
For $\theta \le 1$ equation~\eqref{E:piecewise_field} takes the form:
\begin{equation}
E_s=\begin{cases}
R_n J(\theta-u),& u<\theta\\
0,& u\ge \theta
\end{cases}
\end{equation}
For $\theta >1$
\begin{equation} 
E_s=\f{\rho_s}{d_s}J(1-u); 
\end{equation}
To avoid an unphysical discontinuity at $\theta =1$, we will consider $R_n=\rho_s/d_s$. Then,~\eqref{E:charge_conservation_2}  can be written in a compact form 
\begin{equation}\label{E:charge_conservation_3}
\D{}{x} \left( \lambda^2 \D{u}{x} \right) = 
u - \Gamma \max \left[ 0, \min (\theta, 1)  - u \right],
\end{equation}
where
\begin{equation}
\Gamma =\f{\rho_sd_1}{\rho_1d_s}\gg 1
\end{equation}
and 
\begin{equation}\label{E:current_transfer_length}
\lambda=\left (\f{\bar{R}d_1}{\rho_1}\right )^{1/2}
\end{equation}
is the current transfer length which determines the length scale of the current exchange between the superconductor and stabilizer~\cite{Levin1}.
Taking into account\eqref{E:charge_conservation_1}, the first two terms in the right hand side of~\eqref{E:integrated_internal_heat_sources} take form 
\begin{equation}
\f{\rho_1J^2}{d_1}u^2 +\bar{R}J^2\left (\D{u}{x}\right )^2.
\end{equation}
Taking into account~\eqref{E:charge_conservation_2}, the last term---losses in the superconductor---can be written as follows
\begin{equation}
J_sE_s=J^2(1-u) \left[ \f{\rho_1}{d_1}u-\D{}{x}\left( \bar{R}\D{u}{x}\right) \right].
\end{equation}

We will express the distances in units of thermal diffusion length $l_T$ and time in units of $\gamma^{-1}$, where
\begin{equation}\label{E:thermal_diffusion_length}
l_T =(D_T/\gamma)^{1/2};\;\;\gamma =\rho_1 J^2/d_1C\Delta T.
\end{equation}
Here $D_T = K/C$ is the effective thermal diffusivity of the conductor, $\Delta T\equiv T_c-T_1$, and the increment $\gamma $ determines the characteristic time required for the Joule heat generated in the stabilizer to warm the conductor by the temperature $\Delta T$.

In dimensionless variables, \eqref{E:heat_equation_1} and~\eqref{E:charge_conservation_3} take the form
\begin{align}
&
\begin{aligned}
&\D{\theta}{\tau}-\D{^2 \theta}{\xi^2} = u+r\left(\D{u}{\xi}\right)^2\\ 
&\qquad\qquad{}-(1-u)\D{}{\xi}\left ( r\D{u}{\xi}\right ) -\kappa (\theta -\theta_0)
\end{aligned}
\label{E:heat_equation_2}\\
&\D{}{\xi} \left( r \D{u}{\xi} \right) = u - \Gamma \max \left[ 0, \min (\theta, 1)  - u\right].
\label{E:charge_conservation_4}
\end{align}
with $\tau =\gamma t$ and $\xi =x/l_T$.
Here 
\begin{equation}
\kappa=\f{2K_0\Delta Td_1}{\rho_1 J^2};\;\theta_0 =(T_0-T_1)/(T_c-T_1)<0.
\end{equation}
Notice that~\eqref{E:heat_equation_2} does not depend on the specific form of the constituent relationship between electric field and current density in the superconductor. The specifics of the constituent relationship enters only in the charge conservation condition given by~\eqref{E:charge_conservation_2} and its dimensionless versions~\eqref{E:charge_conservation_3} and~\eqref{E:charge_conservation_4}.

The relative role of the interface resistance is determined by the parameter 
\begin{equation}\label{E:interface_resistance_parameter}
r=\f{\lambda^2}{l_T^2}=\f{\bar{R}}{R_0};\;\;R_0=\f{\rho_1 l_T^2}{d_1}=\f{K\Delta T}{J^2}.
\end{equation}
The results will not depend on the value of $\Gamma $ as long as $\Gamma\gg 1$. Hereafter, for the purpose of numerical calculations,  we take $\Gamma =10^2$. 

\section{\label{sec:uniform_temperature}Results\protect}
As the first step let us consider the interplay between the heat source and cooling power for uniform temperature ($\p^2\theta /\p\xi^2=\p\theta /\p\xi =0$). The analytical solution $u(\theta )$ of~\eqref{E:charge_conservation_4} can be used in~\eqref{E:heat_equation_2} to determine the heat source as a function of temperature. The result is practically the same (as long as $\Gamma\gg 1$) as that shown in Figure~3 in~\cite{PRE}. The system is bistable when $\kappa <\kappa_c=1/(1+|\theta_0 |)$. 
In this case there are two stable uniform modes of operation. One is high temperature $\theta_\text{max}=\theta_0+\kappa^{-1}$ and high dissipation and the other with low temperature $\theta_\text{min}=\theta_0$ and zero dissipation. When $\kappa >\kappa_c$ (cryostable condition), there is only one stable uniform state---the low temperature state with temperature $\theta_0$. 

The system of equations~\eqref{E:heat_equation_2} and~\eqref{E:charge_conservation_4} were solved numerically by using an IMEX Crank-Nicolson/Adams-Bashforth method in conjunction with a fixed-point method to solve the Poisson equation with nonlinear source term.  The purpose of the numerical solutions is to determine the regions of the physical parameters $\{r,\kappa,\theta_0\}$ that correspond to different types of conductor response to initial perturbation, such as the normal zone propagation or formation of the dissipative structures. We also determine the speed of NZP and the margins of stability as the functions of the contact resistance.

The solutions $\theta (\xi, \tau )$ presented below correspond to periodic boundary conditions and the initial condition in the form of a Gaussian in the center of the conductor 
\begin{equation}\label{E:initial_conditions}
\theta (\xi, 0)=(a-\theta_0)\operatorname{e}^{-\xi^2/2\delta^2}+\theta_0. 
\end{equation}
For~\eqref{E:charge_conservation_4} we used the following boundary conditions:
\begin{equation}
u(L)=u(-L)=1.
\end{equation}
This condition means that the current is injected at the ends of the conductor into the stabilizer, as is the case in real experiments. It is important to emphasize that this boundary condition requires that the contact resistance $\bar{R}$ was very low at the ends of the conductor. This allows the current injected into the stabilizer to transit into the superconductor over a very short distance without generating much heat at the ends. Without a low contact resistance at the ends the conductor will become unstable when contact resistance exceeds a certain level. In numerical calculations we have used $\bar{R}(x)$ and, correspondingly, $r(\xi )$ that are constants everywhere, except at the ends of the conductor near $\xi=\pm L$ where $r(\xi )\ll 1$.

\subsection{\label{sec:NZP} Speed of normal zone propagation}
\begin{figure}[ht]
\centering
\begin{tabular}{ll}
\raisebox{0.75\columnwidth}{\parbox{0em}{(a)}}&
\beginpgfgraphicnamed{fig-contours1}
\begin{tikzpicture}
\begin{axis}[enlargelimits=false,axis on top,
width=.90\columnwidth,
height=0.75\columnwidth,
axis y line=left,
axis x line=bottom,
every outer y axis line/.append style={-},
every outer x axis line/.append style={-},
xmin=-50,xmax=50,ymin=0,ymax=80,
xlabel={distance ($x/l_T$)},
ylabel={time ($\gamma t$)},
ylabel style={yshift=-1.2em}]
\ifthenelse{\equal{\publishtype}{electronic}}{
\addplot graphics[xmin=-50,xmax=50,ymin=0,ymax=80]{color_r1k2};
\node (a) at (axis cs:-45,8) {\rlap{\textcolor{white}{$\lambda/l_T=1$}}};}{
\addplot graphics[xmin=-50,xmax=50,ymin=0,ymax=80]{bw_r1k2};
\node (a) at (axis cs:-45,8) {\rlap{\textcolor{black}{$\lambda/l_T=1$}}};}
\end{axis}
\begin{axis}[enlargelimits=false,axis on top,
width=.90\columnwidth,
height=0.75\columnwidth,
axis y line=right,
axis x line=top,
ylabel style={yshift=.4em,rotate=180},
every outer y axis line/.append style={-},
every outer x axis line/.append style={-},
xmin=-20,xmax=20,ymin=0,ymax=12,
xlabel={distance (\centi\meter)},
ylabel={time (\second)},
ylabel style={yshift=-.8em}]\end{axis}
\end{tikzpicture}
\endpgfgraphicnamed
\\
\raisebox{0.75\columnwidth}{\parbox{0em}{(b)}}&%
\beginpgfgraphicnamed{fig-contours2}
\begin{tikzpicture}
\begin{axis}[enlargelimits=false,axis on top,
axis y line=left,
every outer y axis line/.append style={-},
width=.90\columnwidth,
height=0.75\columnwidth,
xmin=-50,xmax=50,ymin=0,ymax=80,
xlabel={distance ($x/l_T$)},
ylabel={time ($\gamma t$)},
ylabel style={yshift=-1.2em}]
\ifthenelse{\equal{\publishtype}{electronic}}{
\addplot graphics[xmin=-50,xmax=50,ymin=0,ymax=80]{color_r4k2};
\node (a) at (axis cs:-45,8) {\rlap{\textcolor{white}{$\lambda/l_T=2$}}};}{
\addplot graphics[xmin=-50,xmax=50,ymin=0,ymax=80]{bw_r4k2};
\node (a) at (axis cs:-45,8) {\rlap{\textcolor{black}{$\lambda/l_T=2$}}};}
\end{axis}
\begin{axis}[enlargelimits=false,axis on top,
width=.90\columnwidth,
height=0.75\columnwidth,
axis y line=right,
axis x line=none,
ylabel style={yshift=.4em,rotate=180},
every outer y axis line/.append style={-},
every outer x axis line/.append style={-},
xmin=-125,xmax=125,
ymin=0,ymax=12,
ylabel={time (\second)},
ylabel style={yshift=-.8em}]\end{axis}
\end{tikzpicture}
\endpgfgraphicnamed
\end{tabular}

\caption{\label{fig:contours}%
Solutions of~\eqref{E:heat_equation_2} and~\eqref{E:charge_conservation_4} for two different values of the contact resistance. Dark (red) color indicates elevated temperature.  The normal zone propagates with constant speed determined by the slope $dx/dt$ at $T=T_c$ ($\theta =1$). The time scale here $0\le\gamma t \le 80$ and the length of the sample $-50 \le x/l_T\le 50$. The values of the contact resistances defined by the parameter $r^{1/2}\equiv \lambda /l_T$ are indicated. The upper scale and the scale on the right show the distance along the conductor in centimeters and elapsed time in seconds, respectively. These values correspond to the specific set of material and operating parameters defined as an example in the text ($l_T=\unit{0.4}{\centi\meter}$ and $\gamma =\unit{6.7}{\reciprocal\second}$).}
\end{figure}
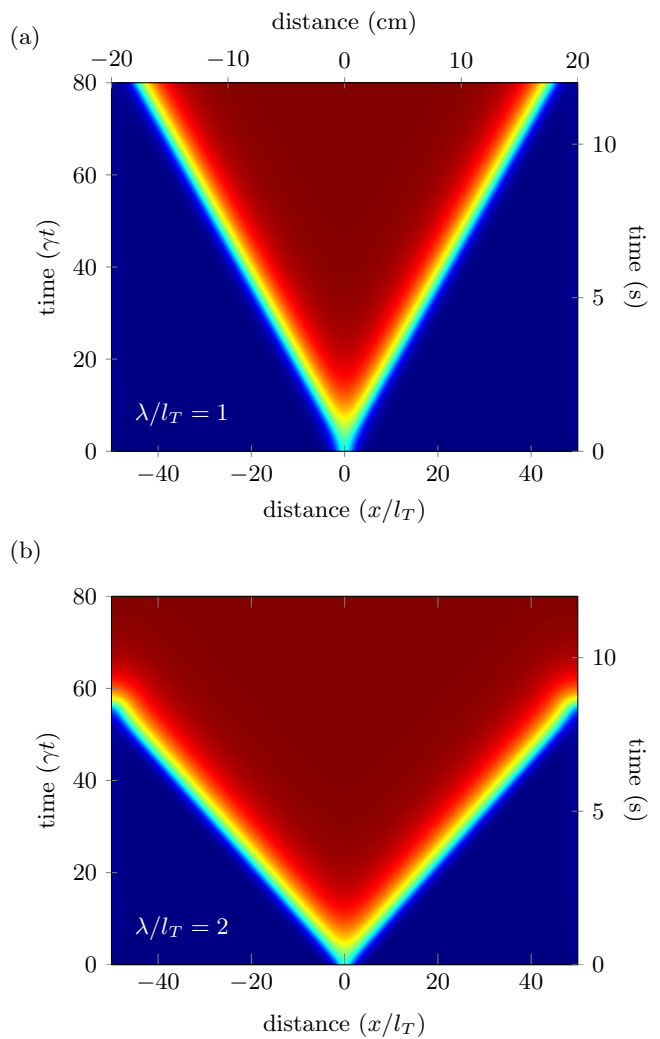%
Figure~\ref{fig:contours} illustrates the growth of the normal zone propagation, which emerges from the initial Gaussian profile.   The figure illustrates the differences between the solutions that correspond to normal zone propagation at different values of the contact resistance.  In each case the cooling constant $\kappa = 0.2$ and the operating temperature $\theta_0=-1$. According to~\eqref{E:dimensionless_linear_temperature_dependence} this corresponds to the conductor operating at $50\%$ capacity, namely $J_c^{0}\equiv J_c(T_0)=2 J$.  The propagation speed noticeably increases with increasing $\lambda$. It should be noted, that for the finite values of the cooling constant $\kappa$ the temperature behind the propagating front does not depend on the contact resistance. In the adiabatic case the situation is different and is discussed in the next subsection.

\begin{figure}[ht]
\centering
\beginpgfgraphicnamed{fig-NZP_speed}
\input{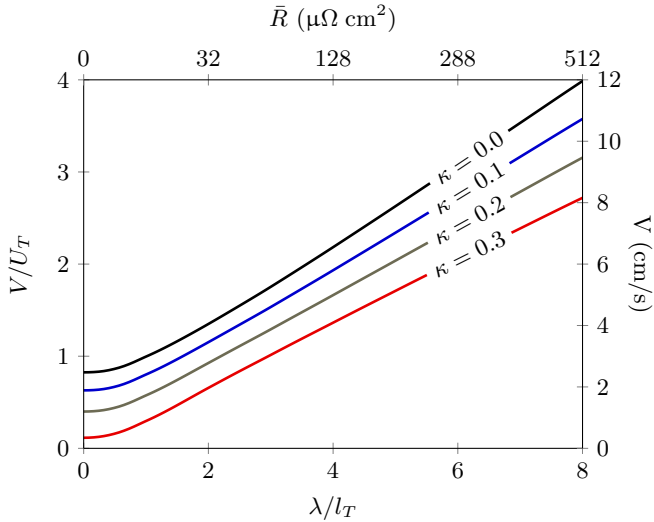}
\endpgfgraphicnamed 
		\caption{	\label{fig:NZP_speed}%
The NZP speed in units of $U_T$ as a function of contact resistance ($\lambda/l_T=(\bar{R}/R_0)^{1/2}$) for several values of the cooling constant $\kappa$. For illustration purposes the scale on the right and the upper scale show the NZP speed and the values of the contact resistance for a set of material parameters described in the text ($U_T=\unit{3}{\centi\meter\per\second}$, $R_0=\unit{8}{\micro\ohm\usk\centi\meter\squared}$).}
\end{figure}
The natural unit for NZP speed is \begin{equation}	\label{E:NZP_speed}
U_T=l_T\gamma =\left (\f{\rho_1J^2K}{d_1C^2\Delta T}\right )^{1/2}.	
\end{equation}
In Figure~\ref{fig:NZP_speed} the speed of normal zone propagation expressed in units of $U_T$ is shown as a function of contact resistance for several values of the cooling constant. For $\lambda/l_T >2$ the speed scales approximately with $\lambda$ which means that the NZP propagation speed is determined by the greater of the two length scales specific to this problem:
\be
V\propto \max[l_T,\lambda]\gamma .
\ee

Let us to flesh out these conclusions using the values of the material parameters representative of coated conductors~\cite{Ekin}. For copper stabilizer we will take $K_1\approx{}${\unit{(4\to5)}{\watt\per\centi\meter\usk\kelvin}};  $C_1  \approx  \unit{1.7}{\joule\per\centi\meter\cubed\usk\kelvin}$;  $\rho_1\approx  \unit{0.2\times 10^{-6}}{\ohm\usk\centi\meter}$,  and $d_1 = \unit{40}{\micro\meter}$ . For Hastelloy substrate we take 
$K_2\approx  \unit{7\times 10^{-2}}{\watt\per\centi\meter\usk\kelvin}$;  $C_2  \approx  \unit{1.4}{\joule\per\centi\meter\cubed\usk\kelvin}$,  and $d_2 = \unit{50}{\micro\meter}$.
Let us also take the critical temperature $T_c\approx  \unit{90}{\kelvin}$ and the operating temperature $T_0=\unit{65}{\kelvin}$.   A reasonable self field value of the critical current density  $J_c (T_0) = \unit{300}{\ampere\per\centi\meter}$,  and the transport current density $J =\unit{150}{\ampere\per\centi\meter}$  (the corresponding value of $ \theta_0 = -1$) . Then, the current sharing temperature $T_1= \unit{77.5}{\kelvin}$ and $\Delta T=T_c-T_1= \unit{12.5}{\kelvin}$.  

The effective  thermal conductance of the wire is dominated by the copper stabilizer
\be 
K=K_1d_1+K_2d_2\approx \text{\unit{(1.6\to2)\times 10^{-2}}{\watt\per\kelvin}}.
\ee
The combined  heat capacity
\be
C=C_1d_1+C_2d_2\approx \unit{1.4\times 10^{-2}}{\joule\per\centi\meter\squared\usk\kelvin}.
\ee
The corresponding thermal diffusivity and the increment 
\be 
D_T\approx \unit{(1.14\to1.4)}{\centi\meter\squared\per\second}; 
\;\gamma =\f{\rho_1 J^2}{d_1C\Delta T}\approx \unit{6.7}{\reciprocal\second}.
\ee
The length scale in this problem is determined by the thermal diffusion length~\eqref{E:thermal_diffusion_length},
\be
l_T\approx \text{\unit{(0.4\to0.45)}{\centi\meter}}.
\ee
The natural scale of the propagation speed~\eqref{E:NZP_speed},
\be
U_T\approx \unit{3}{\centi\meter\per\second}.
\ee
The contact resistance in currently manufactured coated conductors reported in~\cite{Polak}
\be
\bar{R}\approx \unit{5\times 10^{-8}}{\ohm\usk\centi\meter\squared}
\ee
is consistent with other data reported in literature. For this value of the contact resistance the current exchange length $\lambda$~\eqref{E:current_transfer_length},
\be
\lambda\approx \unit{3\times 10^{-2}}{\centi\meter}\ll l_T.
\ee

In the limit of low contact resistance, $\lambda/l_T\ll 1$, the propagation speed $V$ is smaller than $U_T$ in agreement with experimental findings in which the NZP speed was found to be in the range of \unit{(1\to2)}{\centi\meter\per\second}~\cite{Iwasa,Grab,Wang1,H}.  

The characteristic contact resistance defined by the condition $\lambda=l_T$,~\eqref{E:interface_resistance_parameter}, 
\be
R_0=\f{\rho_1 l_T^2}{d_1}\approx \unit{8\times 10^{-6}}{\ohm\usk\centi\meter\squared}.
\ee
Thus, in order to achieve a substantial increase in normal zone propagation speed, the contact resistance has to be increased well over $\unit{10}{\micro\ohm\usk\centi\meter\squared}$. 

\begin{figure}[ht]
\centering
\begin{tabular}{ll}
\raisebox{0.75\columnwidth}{\parbox{-2em}{(a)}}&%
\hspace{-1em}%
\beginpgfgraphicnamed{fig-adiabatic_lines}
\input{adiabatic_lines}
\endpgfgraphicnamed\\
\raisebox{0.75\columnwidth}{\parbox{-2em}{(b)}}&%
\hspace{-1em}%
\beginpgfgraphicnamed{fig-adiabatic_temperature}
\input{adiabatic_temperature}
\endpgfgraphicnamed
\end{tabular}
	\caption{\label{fig:height} (a) The maximum temperature $\theta_{\text{max}}$ inside NZ as a function of its length $\xi_n$ in the adiabatic case ($\kappa=0$) for three different values of the contact resistance. The scale on the right shows the values of $T_\text{max}=T_1+\theta_\text{max}\Delta T$ with $T_1=\unit{77.5}{\kelvin}$ and $\Delta T=\unit{12.5}{\kelvin}$. The top scale shows the length of the NZ in centimeters ($l_T=\unit{0.4}{\centi\meter}$). (b) The rate of temperature increase $d\theta_\text{max}/d\xi_n$ versus contact resistance. The scale on the right shows $dT_\text{max}/dl_n= d\theta_\text{max}/d\xi_n (\Delta T/l_T)$. The upper scale shows the values of the contact resistance corresponding to $R_0=\unit{8}{\micro\ohm\usk\centi\meter\squared}$ in~\eqref{E:interface_resistance_parameter}.}
\end{figure}

\subsection{\label{sec:1d_problem} Temperature rise inside normal zone}

The main danger of slowly propagating normal zone is that it may remain undetected for an extended period of time during which the temperature inside NZ will rise above the damage threshold. This is especially true for adiabatic or near adiabatic conditions. The voltage drop across the normal zone and, correspondingly, the probability of its detection increases approximately in proportion to the length of NZ. Therefore, in order to improve the quench protection quality of the conductor the rate at which the temperature inside NZ rises with the length of NZ has to be made smaller. Obviously, the increasing speed of NZP accomplishes just that. However, the increased contact resistant also increases the amount of power dissipation at the front of the propagating NZ. This makes it necessary to examine closely how the maximum temperature inside the normal zone changes with its length in the worst case scenario of adiabatic NZP ($\kappa =0$).

Figure~\ref{fig:height}(a) shows how the peak temperature changes with the length of NZ $l_n$ defined as the length of a section with $T>T_c$. In adiabatic conditions the temperature inside NZ increases practically linearly with its length. One can clearly see the benefit of increased contact resistance. When the length of NZ reaches $\unit{40}{\centi\meter}$ ($l_n/l_T=100$) the peak temperature of the hot spot may reach $\approx \unit{750}{\kelvin}$ in the  conductor with low contact resistant $\lambda/l_T\ll 1$. In the conductors with substantially larger contact resistance, such that  $\lambda/l_T=4\to 8$, the peak temperature is substantially lower for the same length of NZ.  Of course, the calculated values of temperature here are given only for the purpose of comparison because we do not take into account the changes with temperature in resistivity and other material parameters. 

A more detailed picture of the effect that the contact resistance has on the temperature inside NZ is given in Figure~\ref{fig:height}(b). The figure shows the rate $dT_{\text{max}} /dl_n$ at which the peak temperature increases with the length of NZ (the slope of the curves in Figure~\ref{fig:height}(a)) as a function of the contact resistance expressed as $\lambda/l_T$. This rate allows us to estimate $T_{\text{max}}$ for an arbitrary length of the NZ at a given value of the contact resistance. For $\lambda/l_T\ll 1$ the temperature of the hot spot increases approximately by \unit{18.5}{\kelvin\per\centi\meter} of the NZ length. In a conductor with $\lambda/l_T\approx  8$, the rate of the temperature increase is less than $\unit{5}{\kelvin\per\centi\meter}$.  The inset to Figure~\ref{fig:height}(b) shows the consecutive temperature profiles of the adiabatic NZP for $\lambda/l_T = 4$.

\subsection{\label{sec:stability_margins} Stability margins}
\begin{figure}[ht]
\centering
\beginpgfgraphicnamed{fig-stability_boundaries}
\input{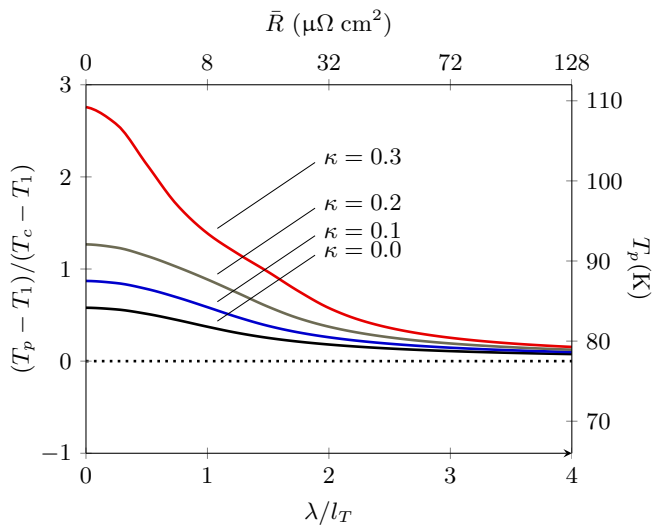}
\endpgfgraphicnamed
	\caption{\label{fig:stability_boundaries}
Stability boundaries as functions of $\lambda/l_T$ for different levels of cooling. The scale on the left shows universal dimensionless units of the peak temperature, Eq. (43).
The scale on the right shows the peak temperature in absolute units for the specific set of material and operational parameters described in the text, $T_p=T_1+a\Delta T$. When the peak temperature of the initial temperature profile exceeds the limit shown by the curves, the normal zone starts to propagate.}
\end{figure}

An ability of a coated conductor to dissipate a certain amount of heat deposited by an external source without triggering normal zone propagation generally declines with increasing contact resistance. We illustrate the reduction in stability margins by using the initial condition in the form of~\eqref{E:initial_conditions} with fixed width $\delta =1.4$ and variable peak temperature $T_p$ determined by the parameter 
\begin{equation}
a=\f{T_p-T_1}{T_c-T_1}.
\end{equation}
Physically, this corresponds to a rapid injection of a certain amount of heat into a small section of the conductor. For a given value of the contact resistance we determine the value of $a$ above which the initial temperature profile gives rise to NZP. For smaller values of the peak temperature the initial profile dissipates without triggering NZP. In Figure~\ref{fig:stability_boundaries} the stability boundaries are shown for several values of the cooling constant. The phase space  $\{T_p,\lambda /l_T \}$ below the respective curve  corresponds to the range of stability, perturbations above the curve are unstable. For illustration purpose the scale on the right shows the peak temperature in absolute units for the set of material and operating parameters described above ($T_0=\unit{65}{\kelvin},T_1=\unit{77.5}{\kelvin}, T_c=\unit{90}{\kelvin}$). 

It is obvious that the stability margins precipitously decline with increasing contact resistance and become less dependent on the cooling conditions. However, our analysis of this model shows that the stability of the initial perturbation depends also on the width of the perturbation.  The greater the width of the initial temperature profile, the more stable the conductor is with respect to that perturbation. When the width of the initial temperature profile is greater or comparable to $\lambda$ its stability is about the same as that of the conductor with $\lambda\approx l_T$. Second, even in the worst case scenario the conductor remains stable as long as the peak temperature is below $T_1$ ($J_c(T_p)>J$) because no current will be diverted into the stabilizer and no heat will be generated (at least within the constituent relationship (11) we have adopted here). For the set of parameters described above the minimum amount of heat per unit area that can be dissipated without triggering NZP can be estimated as 
\be
C(T_1-T_0)\sim \unit{0.17}{\joule\per\centi\meter\squared}.
\ee

\subsection{\label{sec:soliton} Solitons and dissipative structures}
Besides the two uniform modes of operation discussed previously a superconducting current-carrying wire may exhibit a more complex behavior that can be classified as the formation of dissipative structures~\cite{PRE}. The spatially uniform physical systems driven away from thermal equilibrium tend to break the translation symmetry and form steady or time-variable macroscopic spatial patterns~\cite{Cross}. The examples run the gamut from the table-top demonstrations of convection cells, to the sand ripples under water and sand dunes on the ground, to a planetary size phenomenon like the north pole hexagon---a long-lived feature of the atmosphere of Saturn~\cite{S}. 

The spontaneous temperature and critical current modulations along the superconducting wire take place under conditions of strong cooling, above or very near cryostability condition $\kappa >\kappa_c=1/(1+|\theta_0|))$. At this level of cooling the maximum amount of heat generated in the stabilizer is not large enough to maintain temperature above the critical. However, if the contact resistance is large enough, the additional heat generated in the interface by the current passing between the superconductor and stabilizer may be sufficient to sustain a modulated temperature profile.

\begin{figure}[ht]
\centering
\begin{tabular}{ll}
\raisebox{0.75\columnwidth}{\parbox{0em}{(a)}}&%
\beginpgfgraphicnamed{fig-dissipative_structures1}%
\begin{tikzpicture}
\begin{axis}[enlargelimits=false,axis on top,
width=.90\columnwidth,
height=0.75\columnwidth,
axis y line=left,
axis x line=bottom,
every outer y axis line/.append style={-},
every outer x axis line/.append style={-},
xmin=-65,xmax=65,ymin=0,ymax=200,
xlabel={distance ($x/l_T$)},
ylabel={time ($\gamma t$)},
ylabel style={yshift=-0.8em}, 
]
\ifthenelse{\equal{\publishtype}{electronic}}{
\addplot graphics[xmin=-65,xmax=65,ymin=0,ymax=200]{color_r25k6};
\addplot[color=white,dotted,line width=1.0pt] coordinates{(-65,150)(65,150)};}{
\addplot graphics[xmin=-65,xmax=65,ymin=0,ymax=200]{bw_r25k6};
\addplot[color=black,dotted,line width=1.0pt] coordinates{(-65,150)(65,150)};}
\end{axis}
\begin{axis}[enlargelimits=false,axis on top,
axis y line=right,
axis x line=top,
width=.90\columnwidth,
height=0.75\columnwidth,
ylabel style={yshift=+1.2em,rotate=180},
every outer y axis line/.append style={-},
every outer x axis line/.append style={-},
xmin=-26,xmax=26,ymin=0,ymax=30,
ylabel={time (\second)},
xlabel={distance (\centi\meter)}
]
\end{axis}
\end{tikzpicture}
\endpgfgraphicnamed
\\
\raisebox{0.75\columnwidth}{\parbox{0em}{(b)}}&%
\rlap{%
\beginpgfgraphicnamed{fig-dissipative_structures2}%
\input{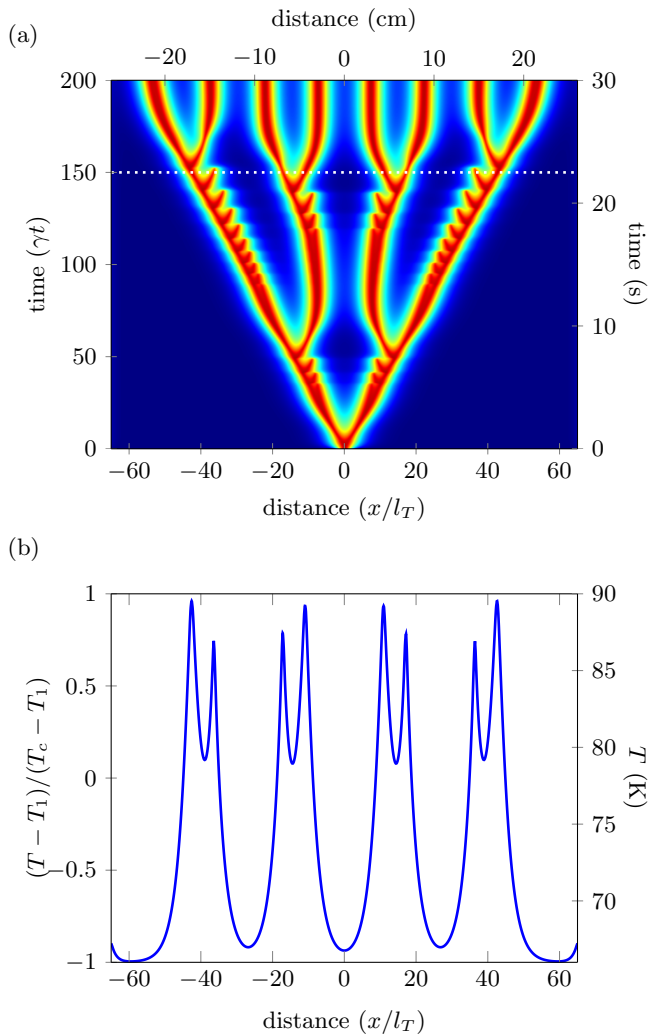}%
\endpgfgraphicnamed}%
\hspace{-.4em}%
\includemovie[poster,autoplay,repeat]{1.005\columnwidth}{.715\columnwidth}{dissipative_structures.swf}%
\end{tabular}
\caption{	\label{fig:dissipative_structures}%
(a) 
The spatio-temporal evolution of a cryostable conductor ($\kappa>\kappa_c$). The dark (red) spots indicate elevated temperatures close to the critical temperature. An individual soliton (hot spot) moves by dividing a single temperature peak into two twins with subsequent dissipation of one of them. The scale on the right and the upper scale correspond to $\gamma=\unit{6.7}{\reciprocal\second}$ and $l_T=\unit{0.4}{\centi\meter}$ (Eqs. (37) and (38)).
(b) The time evolution of the temperature profile. Press `p' to pause/play, `q' to rewind, `f' to advance and 'r' to reverse by a frame. (The animation works in Windows and Mac Adobe Reader.) Alternatively, a snapshot of the temperature profile at $\gamma t = 150$. This is the temperature variation along the dotted line in Figure~\ref{fig:dissipative_structures}(a).}
\end{figure}
A detailed analysis of different scenarios and types of modulation will be published elsewhere.  Figure~\ref{fig:dissipative_structures} presents an example of spatio-temporal development of the initial Gaussian perturbation in cryostable regime, $\kappa = 0.6$, $\theta_0=-1$, and $\lambda/l_T=5.48$ ($\bar{R}/R_0=30$). The color denotes temperatures from the operating temperature (\unit{65}{\kelvin}) to the maximum temperature equal to $T_c$ (\unit{90}{\kelvin}). 

The evolution proceeds as follows: The initial  temperature peak splits into two twin peaks. Each of those peaks, in turn, continue the division, but one of the twins produced in each cycle does not survive. The surviving twin is shifted from the position of its parent and starts the same cycle of division. This results in the directed drift of the soliton  (hot spot) along the wire.  Eventually, the two moving solitons are separated by a distance large enough to even-out the chances of survival of both twins and one of the cycles of division ends successfully, doubling the number of moving hot spots, etc. The peak temperature of the hot spots is close to the critical temperature. Figure~\ref{fig:dissipative_structures}(b) shows the temperature distribution along the conductor at a moment $\gamma t=150$. This profile corresponds to the cut shown by the dotted line in Figure~\ref{fig:dissipative_structures}(a).

It should be noted that a similar type of temperature modulation was discussed in~\cite{Ah1,A,G,Ah2,Gurevich}, where they were called resistive domains. In~\cite{Ah2} an experimental observation of such a resistive domain was reported.

\section{\label{sec:summary}Summary\protect}
There is a viable option to improve the quench protection quality of coated conductors by increasing the contact resistance between the superconducting film and stabilizer. This increases the normal zone propagation speed but decreases the stability margins with respect to localized temperature perturbations. A compromise between these two requirements can be found which may yield a better overall superconducting wire. We should emphasize that the contact resistance here is the resistance to the current exchange between the superconductor and stabilizer, not the resistance to the current flow through the stabilizer itself. It is still desirable to have copper stabilizer of substantial thickness in order to minimize the Joule heat generated in the stabilizer. The contact resistance can be introduced by various means. For example, a very thin film (perhaps \unit{100}{\nano\meter} thick) of highly resistive substance can be deposited on top of YBCO and then covered with protective silver layer and copper stabilizer.

The model presented here describes the process of NZP in a straight conductor where the heat is transferred only along the wire. This condition is similar to that in a superconducting cable. For a pancake coil our conclusions about increasing speed of NZP and reduced stability margins as the functions of the contact resistance remain qualitatively valid. However, a thorough analysis of a pancake coil requires taking into account the heat transfer between the turns, in addition to that along the conductor.  Our results for a pancake coil will be presented elsewhere. 

The practical significance of the dissipative structures and solitons forming in the current-carrying wire under certain conditions is not immediately apparent. However, these are interesting phenomena in their own right and it would be worthwhile to try to observe and study them experimentally. Their detection can be made by conventional electric methods of detection, but a more spectacular result could be obtained with the help of  the real-time magneto-optical imaging~\cite{L,H}.

\end{document}